\documentclass[aps,12pt,tightenlines,amsmath,amssymb,showpacs]{revtex4}
\usepackage{graphicx}
\usepackage{dcolumn}
\usepackage{bm}

\def\Eqref#1{Eq.~(\ref{#1})}
\newcommand{\be}{\begin{equation}}\newcommand{\ee}{\end{equation}}
\newcommand{\bea}{\begin{eqnarray}}\newcommand{\eea}{\end{eqnarray}}
\def\Tr{\mathop{\rm Tr}\nolimits} 
\def\prlsection#1{\bigskip\noindent\textbf{#1}\\ }
\def\adag{a^\dagger}\def\bdag{b^\dagger}
\def\oma{\omega_a}\def\omb{\omega_b}
\def\omA{\omega_A}\def\omB{\omega_B}
\def\c{\cos\theta} \def\s{\sin\theta}
\def\AA{${\cal A}$} \def\BB{${\cal B}$}
\def\Vsb{V_{\mathrm{SB}}}
\def\Vjc{V_{\mathrm{JC}}}
\begin{document}
\title{Ratcheting up energy by means of measurement}
\author{L. S. Schulman}
\affiliation{Physics Department, Clarkson University, Potsdam, New York 13699-5820, USA \vskip-8pt }
\affiliation{\vskip -8pt
Max Planck Institute for the Physics of Complex Systems, N\"othnitzer Str.\ 38, D-01187 Dresden, Germany}
\email{schulman@clarkson.edu}
\author{B. Gaveau}
\affiliation{Laboratoire analyse et physique math\'ematique, 14 avenue F\'elix Faure, 75015 Paris, France}
\email{gaveau@ccr.jussieu.fr}
\date{\today}
\begin{abstract}  
The destruction of quantum coherence can pump energy into a system. For our examples this is paradoxical since the destroyed correlations are ordinarily considered negligible. Mathematically the explanation is straightforward and physically one can identify the degrees of freedom supplying this energy. Nevertheless, the energy input can be calculated without specific reference to those degrees of freedom.
\end{abstract}
\maketitle

\prlsection{Introduction.\label{sec:intro}}
Under some circumstances, a measurement can put energy into a system. This has long been recognized, for example, in locating a quantum particle ``under'' a barrier. However, for the process described below, the decohering of separated systems, one might have thought it to be innocuous, something that takes place without external intervention. We show that in fact the energy increases due to this decohering. Moreover, this has a classical analogue, so that one can say that the energy influx arises because of complimentarity (in a sense to be explained below).

The system is a pair of oscillators, \AA\ and \BB, and could represent an atom and a field or two sorts of oscillators whose self-interactions are substantially harmonic and which couple linearly when close. The excitations are called bosons. We take as Hamiltonian
\be
H=H_0+\Vsb=\oma \adag a + \omb \bdag b + g\left(\adag+a\right)\left(\bdag+b\right)
\label{eq:sbHdef}
\,,
\ee
in the usual notation. This is the spin boson model \cite{note:noshift}. As we proceed we will also refer to recent results on the Jaynes-Cummings model \cite{decoherenceequilibration}, i.e., with coupling $\Vjc=g\left(\adag b+ \bdag a\right)$.

We study the coming together of the two systems for a brief time, followed by their separation (for example, a pulse of light impinging on an atom). Thus one could consider $g$ to be a function of time. Initially we take the density matrix to be a product, $\rho(0)=\rho_{a}(0)\otimes\rho_{b}(0)$; the systems are not entangled. Subsequent to the encounter the state is in general entangled; however, once they are separated the correlations associated with the entanglement can be dropped (provided one does not do an EPR experiment) and the true time-evolved $\rho(t)=\exp(-iHt)\rho(0)\exp(iHt)$ can be replaced by effective, individual density matrices, $\rho_{a}(t)=\Tr_b\rho(t)$ and $\rho_{b}(t)=\Tr_a\rho(t)$.

For both $\Vsb$ and $\Vjc$, diagonalization of $H$ is straightforward. There is however a significant difference that is reflected in the temporal evolution of the operators as well as in the density matrices. For $\Vjc$ there is boson conservation, which implies: 1)~The images of $\adag$ and $\bdag$ under time evolution are linear combinations of the time-0 quantities ($a$ and $b$ do not enter). 2)~If $\rho_{a}(0)$ and $\rho_{b}(0)$ are diagonal in the number operator basis, they stay that way. 3)~On the other hand, coherent states, which are of the form $\exp(z_a\adag+z_b\bdag) |0\rangle$, are mapped into other coherent states (there is a unitary transformation on the $z$'s) and thus preserve non-entanglement. Now for $\Vsb$ there is also a rotation that diagonalizes $H$; after all, as oscillators the coupling is of the form $\gamma x_a x_b$ (where $a=(x_a\sqrt{\omega_a} +ip_a/\sqrt{\omega_a})/\sqrt2$, etc., and $\gamma=\;$const). However, because of the different $\omega$'s that enter for \AA\ and \BB, the time-transformed $\adag$ is a linear combination of \textit{all four} operators, $\adag$, $a$, $\bdag$ and $b$. It follows that the non-entanglement of coherent states is lost, $\rho_a(t)$ and $\rho_b(t)$ are not diagonal in the number (Fock) representation, and the number of bosons is not conserved.

We next consider a more elaborate situation. Imagine a collection of \AA's and \BB's that repeatedly come in contact. Thus \AA$_1$ and \BB$_1$ meet and separate, subsequent to which \AA$_1$ goes on to encounter a different \BB, say \BB$_2$. The usual way to treat the second encounter is to use, for the state of \AA$_1$, the reduced density matrix, $\rho_a(t)$, from its last encounter. If \BB$_2$ also emerged from a similar encounter our estimate of its state is $\rho_b(t)$. So in principle we should maintain a collection of $\rho_a$'s and $\rho_b$'s, and, perhaps randomly, allow pairs to interact. In practice, this procedure does not affect our conclusions and we use the simpler method of taking the output of one encounter, both $\rho_a(t)$ and $\rho_b(t)$, and using it as the input for the next. Thus the initial density matrix for the next encounter is not $\rho(t)$, but $\rho_a(t)\otimes\rho_b(t)$. Based on the assumed multitude of \AA's and \BB's and the unlikelihood of a pair immediately re-encountering one another, this replacement should make no difference.

\begin{figure}
\includegraphics[height=.25\textheight,width=.4\textwidth]{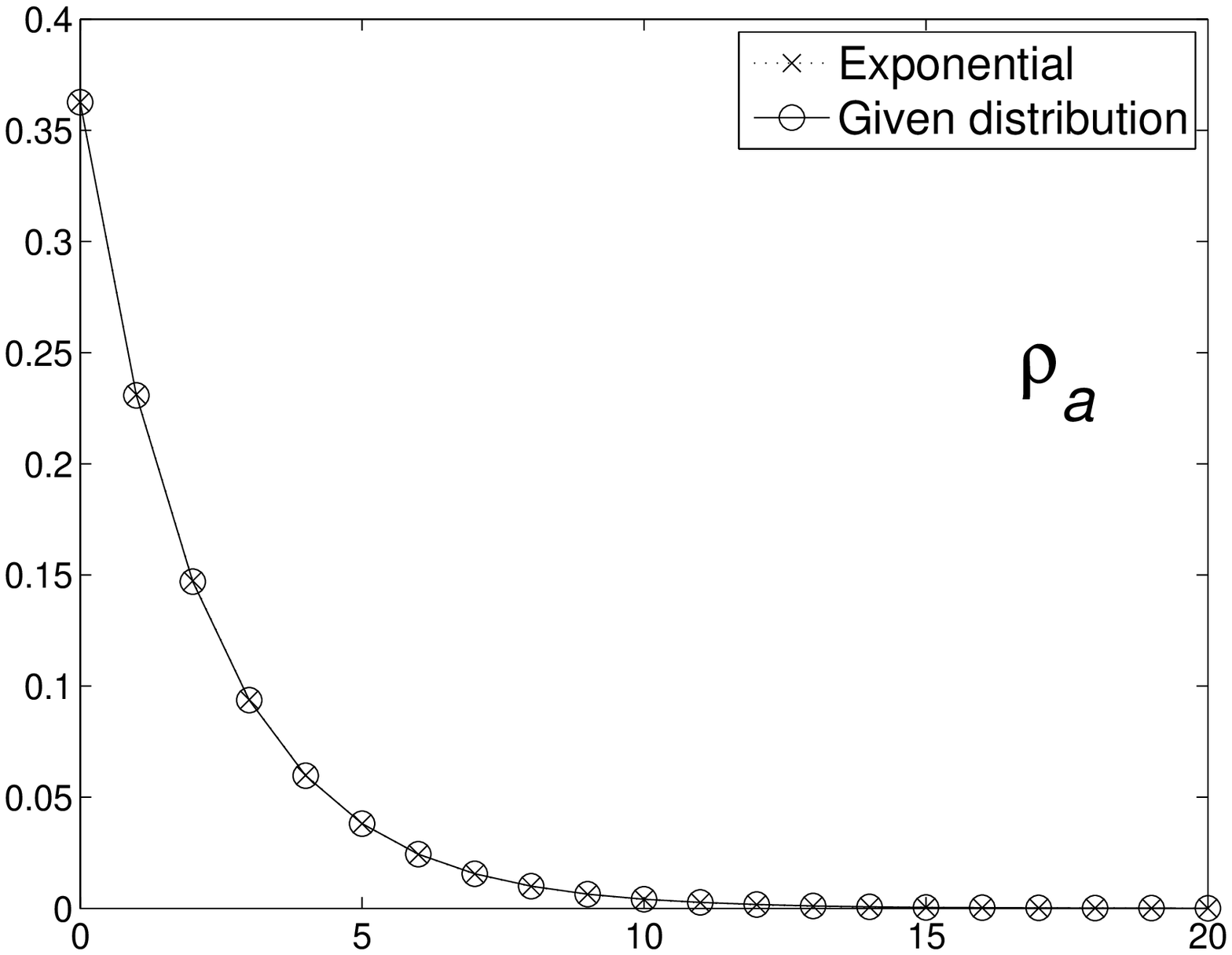} 
\includegraphics[height=.25\textheight,width=.4\textwidth]{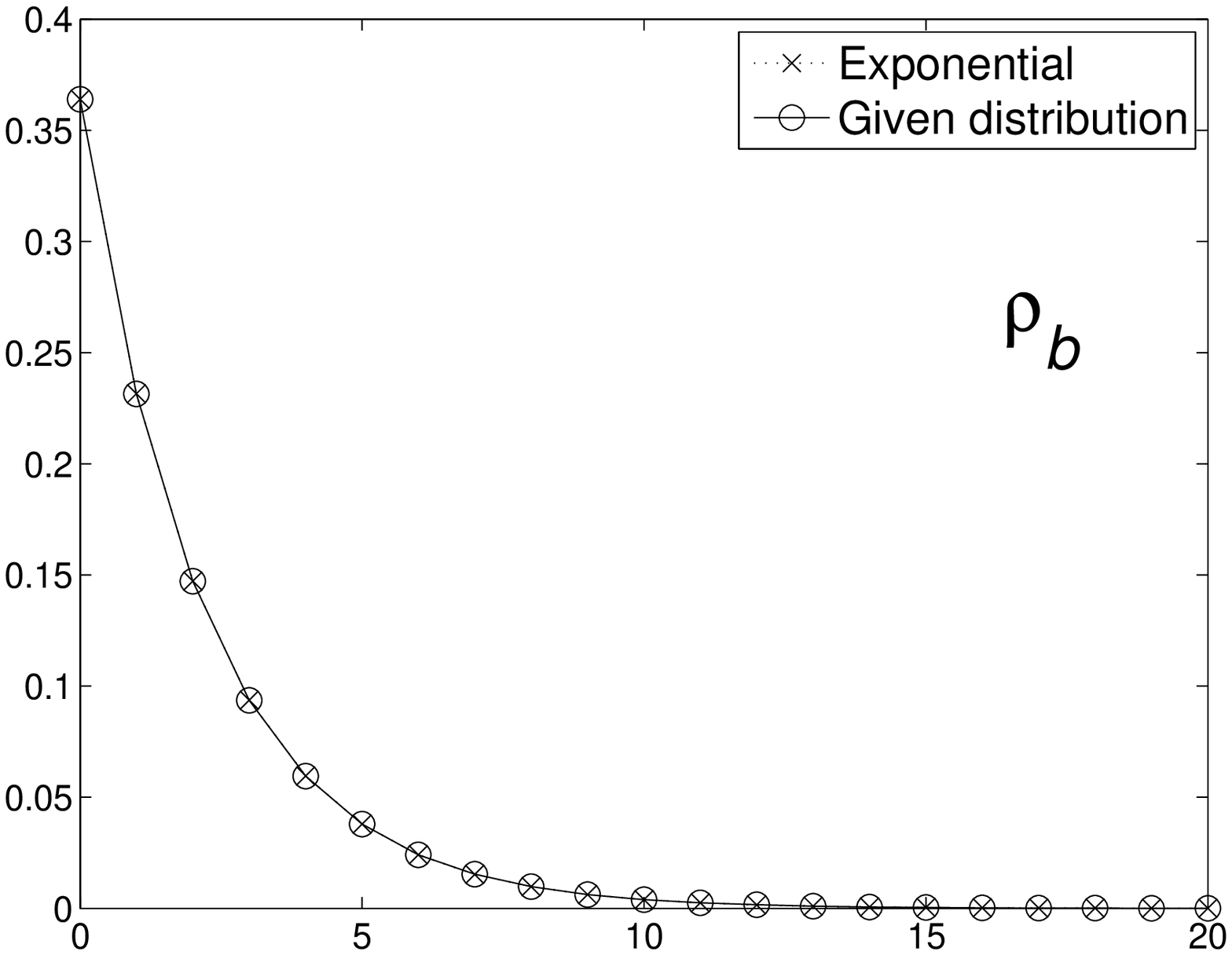}
\caption{Probability distribution of bosons after 25 ``encounters'' for systems \AA\ and \BB\ (marked $\rho_a$ and $\rho_b$). Note that they are essentially the same. \label{fig:spinbosonexponentials}}\end{figure}

\begin{figure}
\includegraphics[height=.25\textheight,width=.4\textwidth]{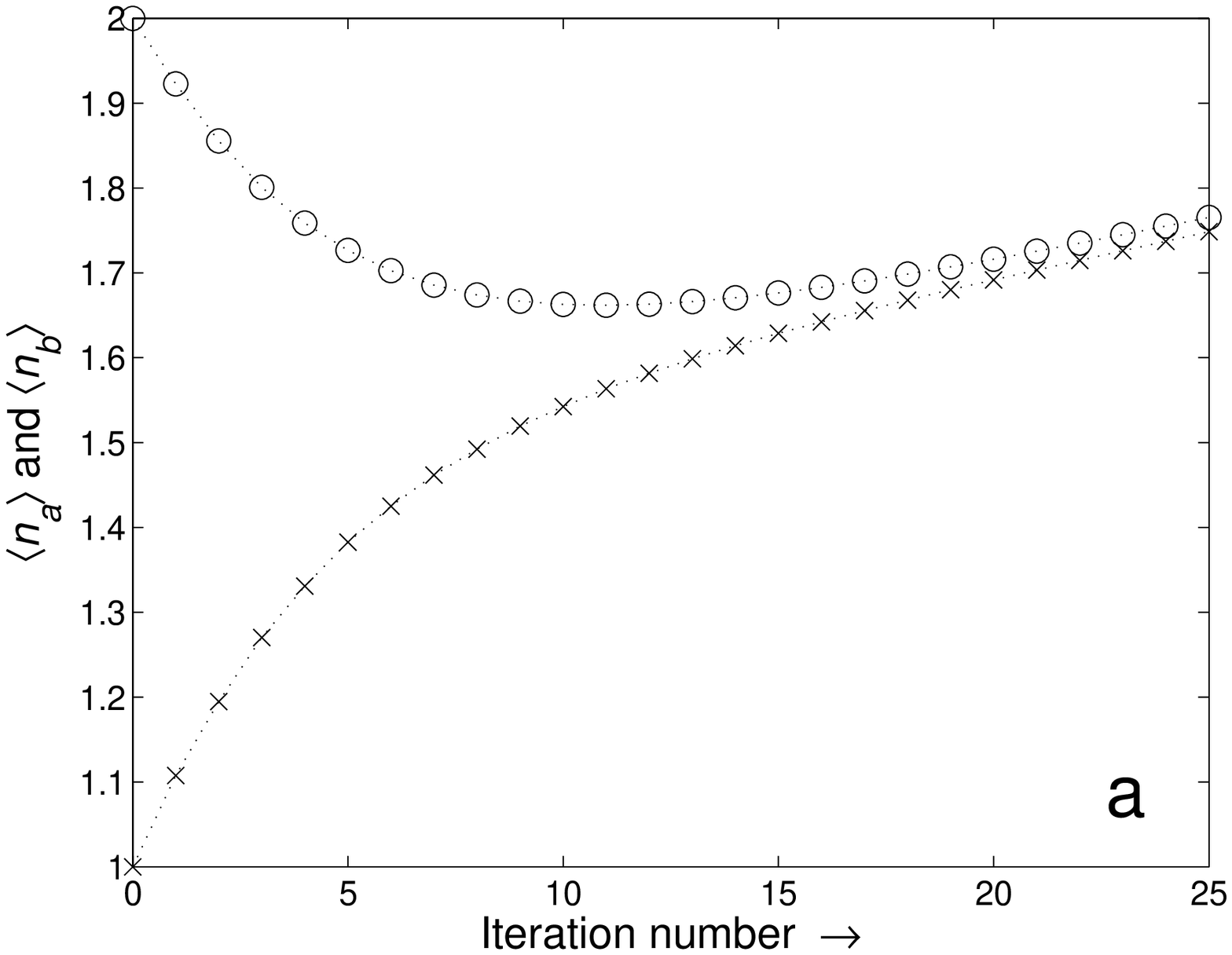}
\includegraphics[height=.25\textheight,width=.4\textwidth]{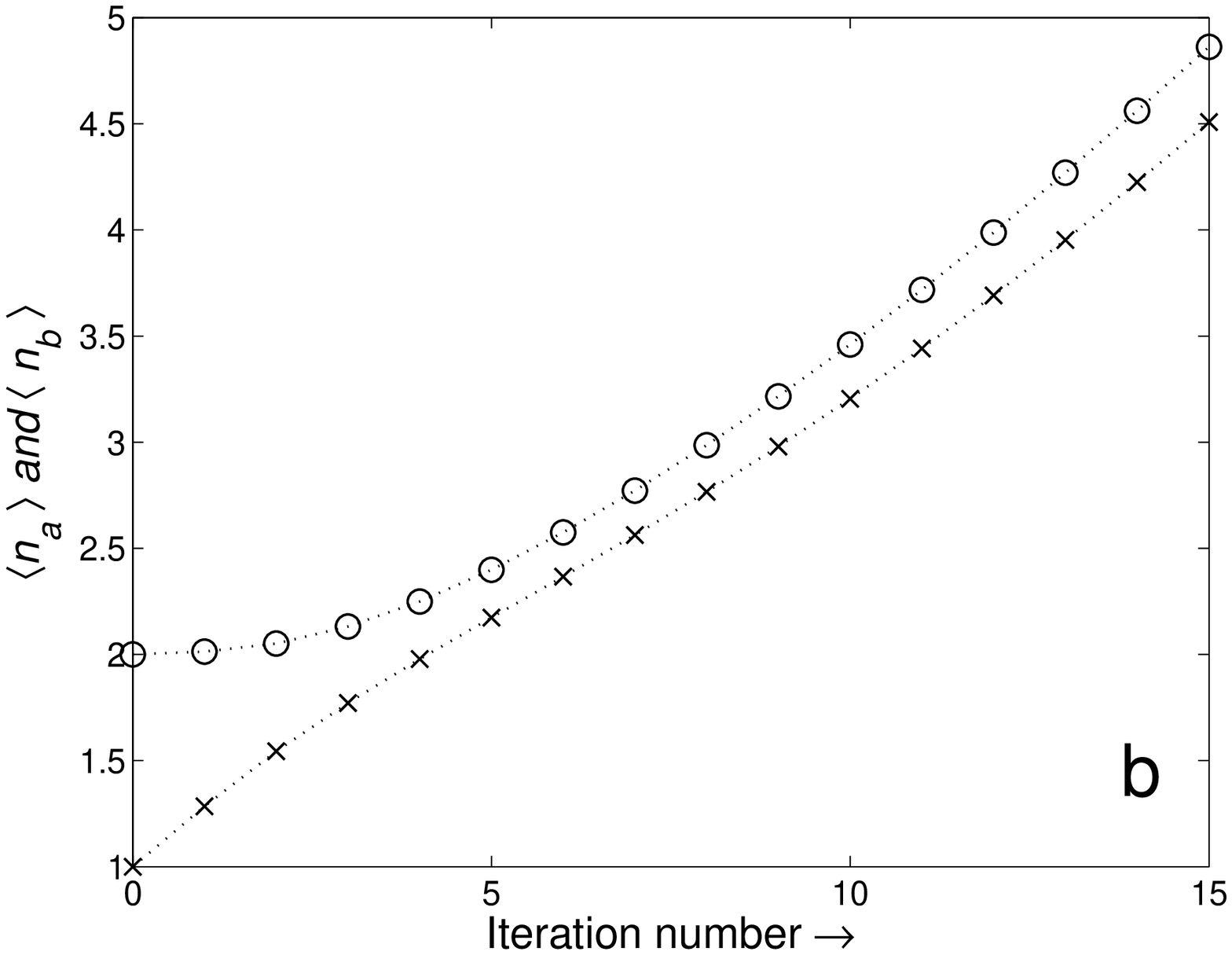}
\caption{Value of the average excitation level in system \AA\ ($\circ$'s) and \BB\ ($\times$'s). Fig.\ \ref{fig:spinbosongrowth}a uses the parameters of Fig.\ \ref{fig:spinbosonexponentials}. For Fig.\ \ref{fig:spinbosongrowth}b slightly different parameters are used, with longer contact times and stronger coupling. In this case $\langle n_a\rangle$ and $\langle n_b\rangle$ grow, but do not tend to a common value.
\label{fig:spinbosongrowth} }\end{figure}

The surprising result is that it makes a big difference. We will explain this both mathematically and physically, but we first present the results. Fig.\ \ref{fig:spinbosonexponentials} shows the probability distributions of boson number after 25 successive encounters, with parameters $\oma=1$, $\omb=2$, $g=0.2$ \cite{note:couplinginstability}, $t=4$, initial states, $n_a=2$, $n_b=1$, and cutoff 20 (i.e., 21 oscillator levels are allowed). The long-time distributions are exponentials with \AA\ and \BB\ having roughly the same dropoff behavior, implying that distinctly different amounts of energy are sequestered in the two modes. For some runs, however, there is a slow drift in the \AA\ and \BB\ values, so that this may be a transient (and numerical considerations prevented our checking). The equality of $\langle n_a\rangle$ and $\langle n_b\rangle$ was observed also for the Jaynes Cummings model \cite{decoherenceequilibration}, and to much greater precision (without the drift that makes us here suspect transience). The remarkable feature of $\Vsb$, not observed for $\Vjc$, is that the average excitation of both oscillators grows as a function of encounter, which is to say, as a function of time. In Fig.\ \ref{fig:spinbosongrowth}a is shown the bosonic content for the same parameters as in Fig.\ \ref{fig:spinbosonexponentials}. In Fig.\ \ref{fig:spinbosongrowth}b is a run with different parameters ($\omb=3$, $g=0.5$, $t=15$). In this case \AA\ and \BB\ seem not to tend to the same limit although both increase. This growth is apparently linear and certainly represents an increase in the total energy in the system. Although one usually thinks of the destruction of quantum correlations as purely an information issue, here it has direct consequences.

\prlsection{Mathematical explanation.\label{sec:mathexplanation}}
Mathematically, we replace a distribution of two variables by its marginals; it's more complicated than in classical probability, since the off-diagonal elements of $\rho$ are complex, but the principle is the same. How this causes trouble can be seen almost without calculation. Let $\psi(0)=|n_a,n_b\rangle={\cal N} {\adag}^{n_a}{\bdag}^{n_b}|0\rangle $, with ${\cal N}=1/\sqrt{n_a!n_b!}$.
Under time evolution the operators become linear combinations of $\adag$, $a$, etc. Thus
\bea
\psi(t)&=& {\cal N}
    \left[\alpha_1\adag +\alpha_2a +\alpha_3\bdag +\alpha_4b\right]^{n_a} \nonumber\\
&&\qquad \times \left[\beta_1\adag +\beta_2a +\beta_3\bdag +\beta_4b\right]^{n_b}|0\rangle \,,
\label{eq:psitFock}
\eea
where $\alpha_k=\alpha_k(t)$ and $\beta_k=\beta_k(t)$, $k=1,\ldots,4$, are explicit functions of time \cite{note:sbdiagonal}. Remarkably, \Eqref{eq:psitFock} implies that under the exact dynamics this state never has more than $n_a+n_b$ bosons of either type. Thinking perturbatively it would seem that more bosons could be created, but this is not so. However, if one takes $\psi(t)\psi^\dagger(t)$ (using \Eqref{eq:psitFock}) and forms from it $\rho_a$ and $\rho_b$, then the correlations of the exact dynamics are lost, and the perturbative considerations apply. Thus the term $\bdag+b$ in the Hamiltonian is applied to $\rho_b$ irrespective of what happened to \AA. This allows unlimited numbers of bosons to be created. (This also lowers $n$, but there is an asymmetry in that $n\geq0$.)

\def\sbdiagonaltext{For $\Vjc$, $H$ is diagonalized by $\tilde a= a\cos\psi+b\sin\psi$, $\tilde b=-a\sin\psi+b\cos\psi$. For $\Vsb$ it's more complicated. The operators diagonalizing $H$ are ${\tilde a}^\dagger = \adag\c\cosh\chi  + a\c\sinh\chi  -\bdag\s\cosh\phi -b\s\sinh\phi$ and ${\tilde b}^\dagger= \adag\s\cosh\phi' + a\s\sinh\phi' +\bdag\c\cosh\chi' +b\c\sinh\chi'$, where $\tan2\theta=4\sqrt{\oma\omb}g/(\omb^2-\oma^2)$, $\omA^2= \frac12(\oma^2+\omb^2) + \frac12(\oma^2-\omb^2) \cos2\theta -g\sin2\theta$, $\omB^2= \frac12(\oma^2+\omb^2) + \frac12(\omb^2-\oma^2) \cos2\theta +g\sin2\theta$, $e^\chi =\sqrt{\omA/\oma}$, $e^\phi=\sqrt{\omA/\omb}$, $e^{\phi'}=\sqrt{\omB/\oma}$, and $e^{\chi'}=\sqrt{\omB/\omb}$. Under time evolution, the new $\adag$ picks up a factor $\exp(-i\omA t)$, $a$ a factor $\exp(i\omA t)$, and similarly for $\bdag$ and $b$. To rewrite in terms of the original operators, invert the transformation, after having applied time evolution, as just described. This gives the functions $\alpha_k(t)$ and $\beta_k(t)$, $1\leq k\leq4$, used in the text}

A second way to see the increase is through short-time behavior. We give an abstract statement of the problem, incidentally showing the phenomenon to be more general. Let
\be
H=H_A+H_B+V_A\otimes V_B
\,,
\label{eq:bgHdef}
\ee
where \AA\ and \BB\ need not be oscillators. At $t=0$ the density matrix is taken to be $\rho(0)=\rho_A(0)\otimes\rho_B(0)$. Under time evolution $\rho(t)=\exp(-iHt)\rho(0)\exp(iHt)$, which is in general entangled; $\rho_A(t)=\Tr_B\rho(t)$ and similarly for \BB. Define $\Delta \rho(t)\equiv \rho_A(t)\otimes\rho_B(t)-\rho(t)$. It is easy to show that
\be
\Delta H\equiv\Tr\left[\Delta \rho(t)H\right]
=
\Tr\left[\Delta \rho(t)\left(V_A\otimes V_B\right)\right]
\,,
\label{eq:bg0}
\ee
which is to say that the expectation of the free Hamiltonian, $H_A+H_B$, is not affected by the replacement. Thus it is the interaction term that gives rise to the effect exhibited in Fig.\ \ref{fig:spinbosongrowth}. For the next step we calculate the short time form of $\rho(t)$, compute its marginals and evaluate the difference. After some calculation (which we will present in \cite{bigratchet}), one obtains
\bea
\Delta H &=&
\frac{t^2}{2}
\Bigl\{
\left[
\Tr_A\left(V_A^2\rho_A(0)\right) -\left(\Tr_A\left(V_A\rho_A(0)\right)\right)^2
\right]    
 \nonumber\\  &&\qquad\qquad \times
\Tr_B\left(\left[V_B,\left[H_B,V_B\right]\right]\rho_B(0)
\right)
  \nonumber\\
&&\qquad +
\left[
\Tr_B\left(V_B^2\rho_B(0)\right) -\left(\Tr_B\left(V_B\rho_B(0)\right)\right)^2
\right]
 \nonumber\\  &&\qquad\qquad \times
\Tr_A\left(\left[V_A,\left[H_A,V_A\right]\right]\rho_A(0)
\right)
\Bigr\}
\nonumber\\
&&\qquad+\hbox{O}(t^3)
\label{eq:expansion}
\,.
\eea
Note that $\left[ \Tr_A\left(V_A^2\rho_A(0)\right) -\left(\Tr_A\left(V_A\rho_A(0)\right)\right)^2 \right]\geq0$, and can only be zero if $\rho_A$ is concentrated on a single value of $V_A$. Therefore to show $\Delta H>0$ we examine the commutators.

The example of interest is when \AA\ and \BB\ are harmonic oscillators. We use a coordinate representation. Thus $H_A=p_A^2/2+\omega_A^2x_A^2/2$, $V_A=\sqrt{|g|}\,x_A$, etc. By direct calculation one finds $[V_A,[H_A,V_A]]=|g|>0$ \cite{note:negativeg}. This shows that the replacement of $\rho$ by $\rho_A\otimes\rho_B$ necessarily increases the expected value of the Hamiltonian, for short times.

\prlsection{Physical explanation.\label{sec:physical}}
The paradoxical aspect of our result is that one expects that in a gas of interacting particles there is little physical significance to their, say, momentum correlations, once they have separated a significant distance. This should apply even more to quantum coherence. It is true that there is a continuing loss of information, but losing quantum correlations should not heat the gas. You don't burn your finger because of a partial trace over a density matrix.

Two phenomena shed light on this situation. First there are the results of entropy production in computing, namely the fact that the one can avoid any thermodynamic cost in a computation, \textit{provided one does not erase} \cite{bennett0}. We will see how this plays a role. Also relevant are physical models of \textit{ratchets} \cite{magnasco, thomas, bier, doering, julicher, gaveau1, gaveau2}, the relation to which will disabuse anyone of the idea that this system could give rise to a perpetuum mobile.

As remarked, the coupling, $g$, can be thought of as a function of time. In fact, it \textit{must} be, since if the coordinates $x_a$ and $x_b$ are physical coordinates then true oscillators will continue to interact at all distances---the farther, the stronger. Alternatively, one could think of these oscillators as internal coordinates on particles with physical position $\bm r(t)$. Then the physical approach and separation of the particles leads to a coupling coefficient of the form $g(\bm r_a(t),\bm r_b(t))$, with $g\to0$ as $|\,\bm r_a(t)-\bm r_b(t)|$ grows. Similar considerations apply if the ``oscillator'' is a mode of a field, although sometimes one can make this idealization without running into trouble~\cite{decoherenceequilibration}. 

Nevertheless, in the situations contemplated here the time-dependence of the coupling constant implies that energy conservation need not apply. For this reason we recall the concept of ratchet, where the turning on and off of a potential induces directional flow. In biological applications this requires external energy, which is the role of \hbox{ATP}. In our case, a full explication will depend on the physical system that \Eqref{eq:sbHdef} represents. Suppose that the oscillator is an internal particle coordinate, borne by the translational degrees of freedom of that particle from place to place, repeatedly encountering other particles. Then as two oscillator-bearing particles approach one another, the coupling energy $g(\bm r_a(t),\bm r_b(t))x_a x_b$ begins to affect the translational motion of the particles themselves. Since the \textit{entire} system can be described by a time-independent Hamiltonian, the source of the energy that enters the oscillator coordinates is necessarily the translational degrees of freedom of the moving particles. If too much is withdrawn, the particles will cease encountering one another (or even separating), and the tracing over the ``other'' degree of freedom inappropriate. 

The fact that the passage of energy to internal degrees of freedom can cool a gas (translationally) is no surprise; what is of interest is that it comes about through the destruction of information, in this case the destruction of quantum coherence. This kind of information loss is not ordinarily considered a source of energy transfer, although, as alluded to above, erasure can have thermal consequences. 

Another argument highlighting the paradoxical nature of our result is that we calculate the energy increase without any explicit model of particle interactions. For the realization mentioned above, $g(\bm r_a(t),\bm r_b(t))x_a x_b$, one would have expected that the dependence of $g$ on distance should play a role in understanding the energy transfer, but somehow we don't need that; decoherence alone drives the process.

\prlsection{Classical oscillators.\label{sec:classicaloscillators}}
To broaden the perspective we mention a classical analogue. Imagine two oscillators with frequencies $\oma$ and $\omb$. At a random time couple them by adding $\gamma x_ax_b$ to the Hamiltonian and allow them to evolve, again for a random time. Repeat the coupling and uncoupling many times. The result is that the energy grows exponentially. This phenomenon is easy to check numerically, but it is also easy to develop a qualitative explanation. The system moves on a 2-dimensional torus in 4-dimensional phase space; at any moment it is on a particular torus. When the coupling is switched (adding or subtracting $\gamma x_a x_b$), the collection of tori changes and the system point continues on the torus associated with the new dynamics. The new torus need \textit{not} have the same energy as the old one. The motion can be described as a jumping from torus to torus, and as a consequence, a random walk in energy. However, because the system is strictly linear, the dynamics is scale-invariant. Therefore the natural description of the random walk is not in energy, but in its logarithm. That a drift-free random walk in $\log E$ leads to exponential growth in $E$ can be seen as follows. Let $u=\log E$ and suppose that the distribution function for $u$ is $p(u)=\exp\left( (u-u_0)^2/2\sigma^2\right) /\sigma\sqrt{2\pi}$. For this $p$, $\langle u\rangle=u_0$. But the expectation of $E$ is not $\exp(u_0)$; rather $\langle E \rangle/\exp\left(\langle \log E\rangle\right)=\exp(\sigma^2/2)$. Under diffusion for time $t$, with diffusion coefficient $D$, the spread, $\sigma^2$, of the distribution changes to $\sigma^2 + 2Dt$.  Since the expectation of the central point $u_0$ does not change, the expectation of $E$ grows like $\exp(Dt)$. As a simpler example on can imagine a single oscillator that switches at random times between two frequencies $\omega$ and $\omega'$. Now the ``jumping'' is between two ellipses in phase space and the energy grows exponentially.

This classical analogue suggests that one can think of the quantum phenomenon as a manifestation of complimentarity. For the quantum system it is not possible to retain all information after a measurement, so that restarting the system with partial information causes it to resemble a classical system for which correlation information has been destroyed ``by hand.'' (In the work just described we accomplish this by letting the system run for ``random'' times. With non-random times it will sometimes grow exponentially, sometimes oscillate.) 

\prlsection{Acknowledgements.}
We thank A. Buchleitner, L. Davidovich, S. Flach, N. Perkins, and M. Scully for helpful discussions. This work was supported by NSF Grant PHY 0555313. 


\begin{thebibliography}{15}
\bibitem[{not({\natexlab{a}})}]{note:noshift}
\bibinfo{note}{The ``counter term'' (cf.\ \cite{weiss}) is not relevant to our
  considerations.}
\bibitem[{not({\natexlab{b}})}]{note:couplinginstability}
\bibinfo{note}{For large enough coupling coefficient (``$g$'' or ``$\gamma$'')
  the potentials become unstable. All our quantum and classical results are for
  smaller coefficient values.}
\end{thebibliography}
\end{document}